# Validation and Transparency in AI systems for pharmacovigilance: a case study applied to the medical literature monitoring of adverse events

Bruno Ohana
Biologit Ltd
bruno.ohana@biologit.com

Jack Sullivan
jack.tv.sullivan@yahoo.com

Nicole Baker
Biologit Ltd
nicole.baker@biologit.com

## Abstract

Recent advances in artificial intelligence applied to biomedical text are opening exciting opportunities for improving pharmacovigilance activities currently burdened by the ever growing volumes of real world data. To fully realize these opportunities, existing regulatory guidance and industry best practices should be taken into consideration in order to increase the overall trustworthiness of the system and enable broader adoption.

In this paper we present a case study on how to operationalize existing guidance for validated AI systems in pharmacovigilance focusing on the specific task of medical literature monitoring (MLM) of adverse events from the scientific literature. We describe an AI system designed with the goal of reducing effort in MLM activities built in close collaboration with subject matter experts and considering guidance for validated systems in pharmacovigilance and AI transparency. In particular we make use of public disclosures as a useful risk control measure to mitigate system misuse and earn user trust.

In addition we present experimental results showing the system can significantly remove screening effort while maintaining high levels of recall (filtering 55% of irrelevant articles, on average, for a target recall of 99% on suspected adverse articles) and provide a robust method for tuning the system's desired recall to suit a particular risk profile.

# 1. Introduction

Medical literature monitoring (MLM) of adverse drug reactions and special situations is the task of periodically searching the scientific literature for potential safety events associated with a product of interest with the goal of reporting findings to regulatory authorities for appropriate action, or gaining insights on a product's safety and efficacy profiles. Periodic MLM is a regulatory requirement for marketed products in many jurisdictions (ex: [1,2]) and also performed by regulators, the European Medicines Agency being an example [13]. This activity is typically performed by a team of experts in a pharmacovigilance team with responsibility for a marketed product.

A typical MLM workflow is shown on Figure 1: a pharmacovigilance specialist will craft a database search based on query strings targeting the product of interest and relevant terms indicative of adverse events or special situations (off-label use, paediatric, pregnant or elderly populations, etc). This search is then run on a number of scientific databases, results are extracted, de-duplicated and undergo initial screening by a specialist based on the title and abstract of the article. Should the article be deemed relevant the process continues by carrying out a deeper inspection based on the full text of the article and subsequent quality control.

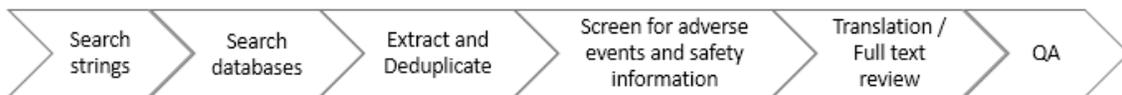

*Figure 1- A typical MLM workflow*

Screening of abstracts will vary according to the task objectives: for periodic surveillance of individual case safety reports (ICSR) specialists look for safety events in articles where identifiable patients can be found. Other screening tasks may look for aggregated safety data from patient studies or target the discovery of possible new safety signals. What is common however is the activity will demand more resources as the input grows in volume. Indeed, this volume continues to grow with more publications being available and more products entering the market every year, increasing the demand for specialist resources in pharmacovigilance teams. At the same time, recent surveys indicate pharmacovigilance workflows still lag in automation adoption [3] presenting an ideal opportunity to consider Artificial Intelligence (AI) solutions to deliver productivity gains and higher quality results.

Reaching this goal requires technology development coupled with adequate process maturity that ensures validated and trustworthy AI systems, attracting interest from regulators and industry bodies who have put forward guidance in developing AI systems for pharmacovigilance [6,7,19]. Leveraging these developments, we describe an AI system for medical literature monitoring of the scientific literature and corresponding development process serving as a case study operationalizing current guidance. In addition we present experimental results demonstrating how the system can realize productivity gains in MLM workflows, and discuss lessons learned and future development opportunities.

## 2. Considerations and Guidance for AI Systems in Pharmacovigilance

Software engineering practices for pharmacovigilance typically follow established computer systems validation (CSV) guidance in the International Society for Pharmaceutical Engineering's Good Automated Manufacturing Practices (ISPE GAMP [5]). To address specific considerations of AI software systems, the guidance proposed by TransCelerate [6], extends CSV leveraging existing regulatory guidance and best practices, in particular FDA's seminal work on AI/ML software as a medical device [7]. Despite the TransCelerate guidance being particularly well suited for AI in pharmacovigilance, this remains an active area where regulatory agencies and industry bodies continue to seek feedback and evolve guidance on the use of AI. We refer readers to a recent survey of ongoing activities on this topic in [19].

We also observe considerations from the AI transparency literature may contribute to the same goals of guidance for AI in pharmacovigilance: from [6], clearly stating a model's *intended use* is an important requirement for validated systems and expected to be part of a system design specification, while the *public* disclosure of an AI system's intended use is encouraged as a transparency measure that makes clear the system applicability and trade-offs, reducing the risk of misuse [14]. AI disclosure considerations and proposals can be seen in the work on model FactSheets [16], Model Cards [20], Datasheets for Datasets [21] and the AboutML initiative [14].

## 3. AI Development Approach

Our AI development approach has the following goals: (a) incorporate existing best practices in computer systems validation and supporting processes for change control, risk management, etc. (b) ensure decisions at important stages of the development process are traceable and can be later interrogated, (c) consider the existing guidance for developing AI systems for pharmacovigilance, notably considerations presented in [6] and finally (d) leverage AI transparency best practices to mitigate risk and improve trustworthiness of the system.

The processes supporting the lifecycle of the AI system are heavily anchored in artifacts maintained in a quality management system (QMS), where documentation changes are fully traceable and undergo established approval processes. This ensures decisions relating to the system design, data sourcing and labelling, model training and testing have a full audit record. Using a QMS also facilitates the integration and reuse of supporting processes for CSV such as change control, risk and documentation management.

The backbone of our process is described in the *AI development lifecycle* document: a multi-stage development process that delivers milestones iteratively through planned releases. Each stage is common in scope to traditional software development, but includes considerations and best practices specific to AI development. Table 1 describes process stages and included deliverables of particular interest to AI.

Deriving directly from existing documentation in the QMS, we include *public disclosures* as part of our release process: technical details of the AI system and models are documented in a factsheet [1] derived from the one proposed in [16]. Corresponding user-facing product documentation is also updated as appropriate, as illustrated in Figure 2.

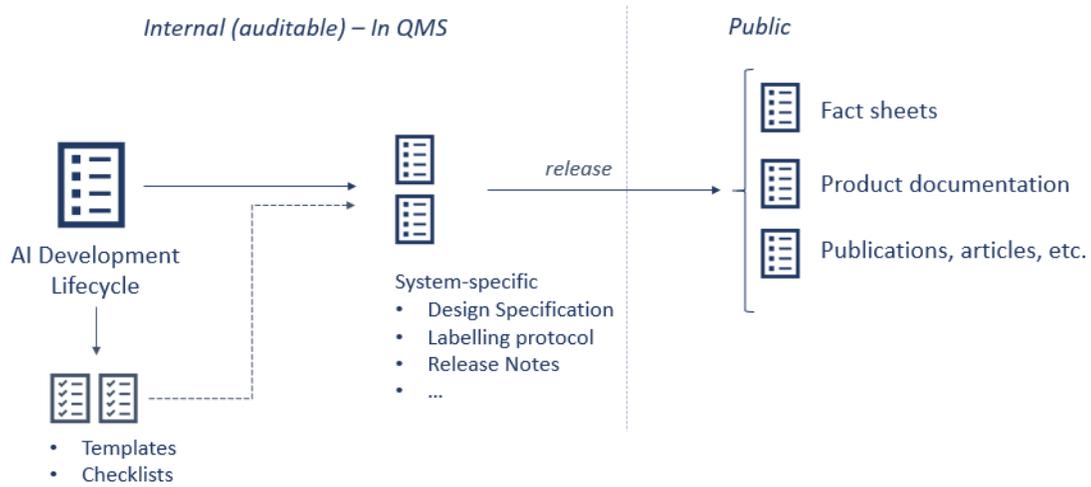

*Figure 2- AI development lifecycle documentation artifacts.*

In addition to documentation the code, data and models must also follow standardized control mechanisms that ensure an auditable record of changes. A combination of git [8] and the open source Data Version Control tool (DVC) [9] provided a suitable toolkit for ensuring there is an auditable record of changes across these artifacts. Git is a widely used distributed version control system providing a permanent change record of code and other artifacts in its repository; DVC follows a "git-like" workflow to accomplish similar goals for data and model artifacts with additional features specific for data science workflows such as running repeatable experiments. By not deviating significantly from typical software engineering workflows this approach makes adoption easier and leverages existing tooling and best practices built by the developer community. Table 1 outlines traceability controls in place across the development lifecycle stages:

---

[1] Companion factsheet to this study is available from: https://github.com/biologit-engineering/factsheets

Table 1 - Traceability mechanisms of the AI development lifecycle process.

| Phases | Traceability |
|---|---|
| Planning<br><br>Requirements<br><br>Design | Documentation artifacts are controlled documents in the QMS. (Key deliverables: AI development lifecycle, release plan, design specifications, data labelling plan and labelling protocol). |
| Development | Model code, artifacts and data are version controlled with git + DVC.<br><br>Changes undergo traceable review and approval process via pull requests. |
| Testing | Unit testing is automated and integrated into the code review process.<br><br>Automated model testing reports (integration testing) produced with every release; results are version controlled.<br><br>Functional testing for computer system validation includes AI-specific features, artifacts are tracked as controlled documents in the QMS. (Key deliverables: user and functional requirements matrix, installation qualification, operational qualification and validation summary report) |
| Release | Releases follow change management process and are traceable via controlled documents; Code releases undergo review and approval process from GitHub. |
| Monitoring | Periodic monitoring is carried out via functional testing and benchmarking experiments. Results are tracked as controlled documentation in the QMS. |

## 4. System Description

The system acts as an additional stage to the original MLM workflow, shown in Figure 3, where model predictions *filter* results retrieved from searching the literature according to whether these may contain relevant safety information.

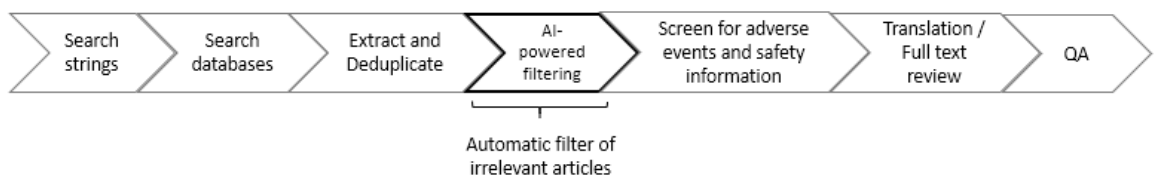

*Figure 3- AI-enabled MLM screening workflow.*

Using the title and abstracts of an article as input the system issues a prediction on whether the article contains one or more *suspect adverse event*. Suspected adverse articles proceed for further screening. To ensure compatibility with as many data sources as possible, no additional article metadata is used by the system other than the text of the title and abstract.

## 4.1 Inference Pipeline

The inference pipeline comprises a pre-processing stage to clean and tokenize the input, detect abstract language and perform entity extraction of patient mentions used in later stages. The model inference stage encodes the normalized text into features and runs the prediction step of the machine learning models, producing raw model predictions. Next, a post-processing rules-based stage produces the final predictions and an explanation step computes metadata to help users interpret model predictions. The design of the pipeline is modular so that each stage can produce additional outputs to be consumed downstream. For instance, additional machine learning models and entity extractors can be added, and later combined at the rule-base stage.

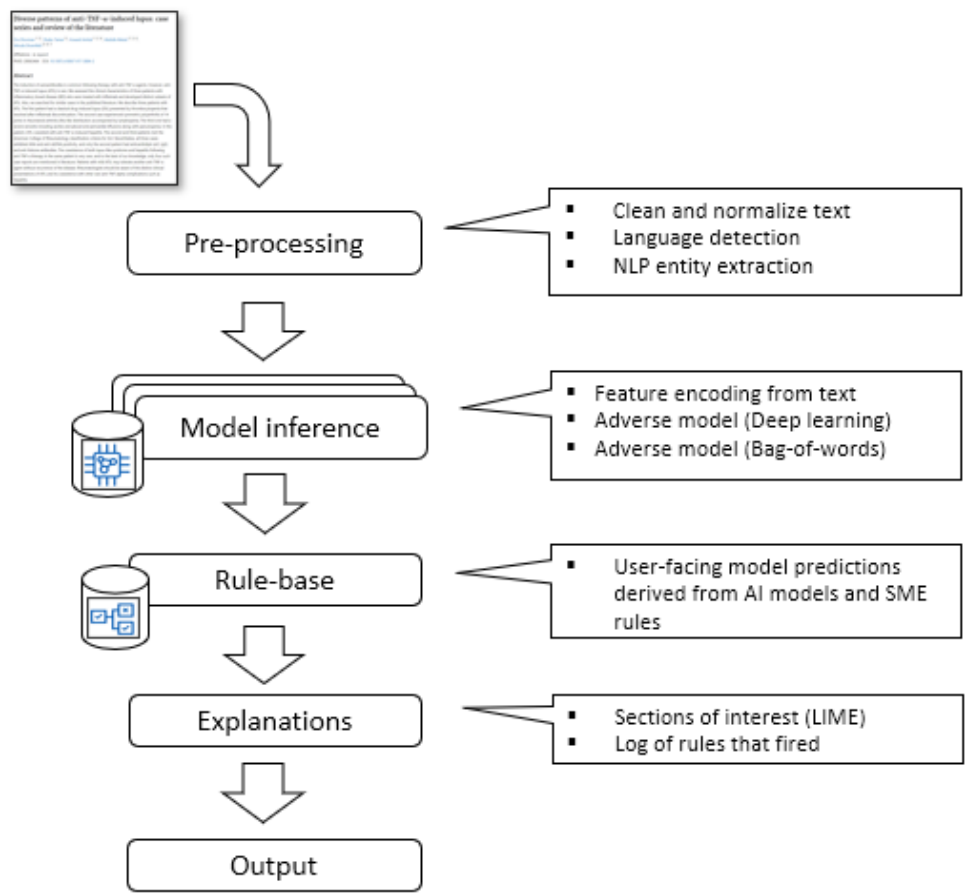

*Figure 4 - Inference pipeline.*

The current implementation uses two machine learning models at the model inference stage: The main adverse detection model is a refinement of our previous work [4] and employs a multi-layer neural network (NN) architecture organized as follows: an initial embedding layer converts tokens into vector representations using a combination of pre-trained word embeddings built with a biomedical text corpus [10] and additional trainable embedding layers derived from part-of-speech tags and dependency parsing tags. The embeddings are combined and processed by a series of convolutional layers followed by a LSTM recurrent layer [17] and an attention layer. Regularization is applied across the network architecture by using drop out during training and the use of batch normalization layers. Complementing the NN model a second, separate model uses bag-of-words (BOW) representation as features from 1-gram and 2-grams and is trained with a random forest estimator.

The rule-based stage of the pipeline takes the output of all the previous stages as input, and computes predictions by evaluating a set of pre-defined rules created in collaboration with pharmacovigilance experts. This approach is used to override model predictions when SMEs deem it is the safer option, or to ensure predictions are only made within a pre-defined operating envelope. In the case when rules override machine learning predictions, the execution log provides an additional layer of explainability to results. Figure 5 outlines the rules implementation for issuing a suspect adverse prediction: input attributes extracted during pre-processing are verified against the model's operating envelope. Currently this comprises of verifying the inferred language of title and abstract is English, the input size is within boundaries seen during training, or input is of a known invalid format (ex: abstract errata descriptions).

The prediction scores from the machine learning models are then checked against thresholds configured as rules. If a prediction using the NN model is not adverse, it is then checked against the BOW model at a threshold level deemed of high confidence. Finally, the system checks if an identifiable patient was found by the entity extraction pre-processing stage (ex: "46-year-old male patient"). This rule reflects pharmacovigilance experts advice in that identifiable patients appearing on case report or case series articles should undergo a closer inspection.

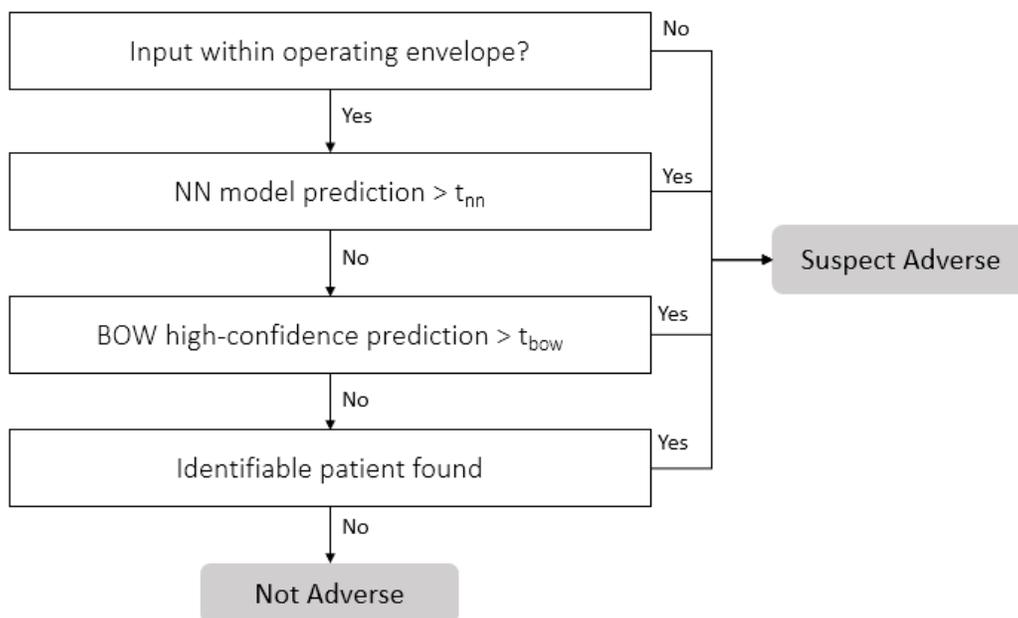

*Figure 5 - Rules implementation.*

At the explanations stage we employ a variation of the LIME technique [11] to extract sections from the abstract most relevant to the predicted class. LIME is a model-agnostic explainability technique that generates predictions in the vicinity of the input by randomly applying small modifications to the input text. Explanations to the model prediction can be derived from the modifications that affected the predicted label most significantly. From experimentation we find the technique is best applied in our case when using entire sentences as the unit of modification as opposed to individual tokens.

## 4.2 Data Labelling

The labelling approach must take into account the fact that abstracts may contain only partial safety information: it is not always possible to fully determine the details of an adverse event from the abstract or title, and further inspection in the article full text may be needed. For the same reason a suspect event may apply to any product being mentioned in the text - ie. it may not be possible to establish causality. Minimizing the risk of filtering relevant articles is also an important consideration, and finally the model should be broadly applicable to other related MLM activities introduced in Section 1.

Training data was curated and labelled in advance by pharmacovigilance specialists and does not dynamically change with user input, thus categorizing the resulting models as "static". Static systems are desirable from a validation standpoint as they align with guidance from [6] and can better follow testing and change management practices for computer systems validation.

The machine learning task is a *binary text classification* problem where the prediction identifies an abstract as *suspect adverse* or not. For data labelling this is defined as an abstract that either contains explicit mentions of an event or implicit mentions that may be fully described in the full text of the article. This definition allows for building classifiers that can be used in many scenarios that need the detection of safety events while at the same time minimizing the risk of false negatives.

The dataset is built from a selection of publicly available abstracts from the biomedical literature. Articles were iteratively selected using pre-defined keyword searches that (1) reflected compounds belonging on one or more categories for therapeutic use from the MeSH taxonomy [12], or (2) were part of the European Medicines Agency list of products under surveillance for medical literature monitoring [13] or (3) were retrieved using commonly used adverse terms or terms for special situations (paediatric, off-label, pregnancy-related terms, etc.). The data labelling and training objectives are recorded in the QMS and form part of a continuous process of model re-training and release. The entire labelling process is the responsibility of pharmacovigilance specialists, and all labelled data undergoes sampled quality checks by a second annotator to ensure labels remain consistent. Inter-annotator agreement is monitored using Coen's Kappa metric, and additional reviews are triggered should it fall below an agreed threshold.

Table 2 - Suspected adverse examples.

| Example Excerpt |
|---|
| "A 58-year-old woman developed hypercalcemia and while on vitamin D supplements". (direct and less ambiguous) |
| "Twenty-two patients experienced peripheral neuropathy, and two had severe neuropathy". (event may be associated to treatment) |
| "Although dose reduction was required in 20% of the patients, no adverse event that led to the discontinuation of treatment was observed." (more ambiguous, implies adverse event led to dose reduction) |

### 4.3 Performance Metric

Mistakes in the prediction of safety events have an asymmetric risk profile: articles falsely identified as suspect adverse (false positive) would incur incremental screening effort, but articles falsely identified as *not* suspect adverse (false negative) negatively impact what safety information is detected. In deciding the performance metric to optimize, the least risky approach is one that minimizes false negatives, even at the expense of additional effort.

The adverse model is parametrized for a desired target recall, ensuring false negatives remain low: with recall fixed at a sufficiently high level, a metric that reflects the additional effort caused by false positives should be minimized. We use the *false positive rate* (fpr), defined as the ratio of false positives (FP) to the number of ground truth negative examples (N) given by:

$$\frac{FP}{N} = \frac{FP}{FP + TN}$$

Where TN is the number of true negatives. Thus, the performance target is the *minimization of false positive rate at a desired target recall*.

### 4.4 Risk and Controls

Framing current guidance considerations as *risks* addressed by one or more *controls* provided a useful mechanism to translate guidance into action items during the development life cycle. This approach is summarized for our system in Table 3. We note that in many cases public disclosures can be used as an effective risk mitigation tool.

Table 3 - AI system risks and controls.

| Category | Risk | Applied Controls |
|---|---|---|
| Intended use | The AI system is applied outside of its intended use and operating envelope. [6, 14] | Inference pipeline verifies input within operating envelope (rules stage).<br><br>Publicly disclose intended use in the system's fact sheet. |
| Intended use | User over-confidence on predictions from the AI system, performance on data not seen in training is not available. [6, 18] | Publicly disclose performance metrics and justification in system fact sheet.<br><br>System supports different levels of supervision to facilitate validation and adoption. |
| Intended use | Full automation derived from model predictions may hide incorrectly predicted documents from users. [6] | System supports diverse levels of supervision and validation, and does not require mandatory full automation.<br><br>All historical results are auditable irrespective of prediction result. |
| Data | Inconsistency in labelling training data may lead to poor performance. [6] | Data labelling performed by pharmacovigilance specialists with MLM background; quality checked and monitored for inter-annotator agreement. |
| Data | Data used in training is not representative for the intended use, or is biased towards certain scenarios, leading to unexpected results. [6] | Training data composition, intended use and domains not covered in the training data are disclosed in the system fact sheet. |
| Data | Input data is not in expected format and may generate spurious predictions. [18] | Inference pipeline performs check of input data for language and operating envelope.<br><br>AI system is integrated into data pipeline where ingested data is from verified sources. |
| Data | Poor quality of training data may affect model performance. [6] | All training data collected is labelled by pharmacovigilance specialists.<br><br>Data labelling process includes quality check and verification of inter-annotator agreement. |
| Model | Model over fitting [6, 18] | NN model trained with regularization and supplemented by additional BOW model.<br><br>Model performance is reported against a left-out test set, and an independent dataset from another distribution. |
| Model | Data leakage into training set may lead to misleading performance data. [6, 18] | Integration testing includes validation against data leakage. |
| Model | Model drift: age/relevancy of training data. [6, 18] | Periodic model releases are envisaged with corresponding updates to public disclosure.<br><br>Training data is regularly updated following AI lifecycle and change management process. |
| Model | Robustness to adversarial attacks. [6] | The model is currently operating on input data sources vetted by the engineering team. |
| Transparency | Model predictions are not easily understood, and less trustworthy to users [6] | Model predictions are accompanied by explanatory statements derived from the rule-based stage; using LIME technique for model explanations. |
| Transparency | Model design and validation decisions are not visible or clearly justified, becoming less trustworthy to users [6, 14, 18] | Model design and validation decisions are publicly disclosed in the system fact sheet; In-depth documentation is tracked in the QMS and auditable upon request. |

## 5. Experimental Results

The training set is built by randomly sampling 90% of all available labels (n=20,531). Sampling is stratified across the labelling categories used to select data discussed in Section 4; the minority label is randomly up-sampled to produce a balanced number of samples for each label.

The remaining 10% of the original dataset is split evenly into a validation set and a test set (each with n=1141). The training set is used to train the core machine learning models, while the validation set is used for hyper parameter tuning. The test set is set aside solely to report model performance results.

Once the BOW and NN models have been trained, a hyper parameter tuning step employs grid search to determine suitable values for model thresholds (Figure 5) that minimize the false positive rate on the adverse class for a target recall. The experiment is repeated 10 times by randomly re-sampling the test/validation sets. Figure 6 shows the trend for the desired target recall (x-axis) vs. obtained recall on test set (y-axis) using hyper parameter search for NN and BOW as standalone models. The box plot shows inter-quartile ranges across all 10 x experiment runs, with minimum and maximum values extending from the box.

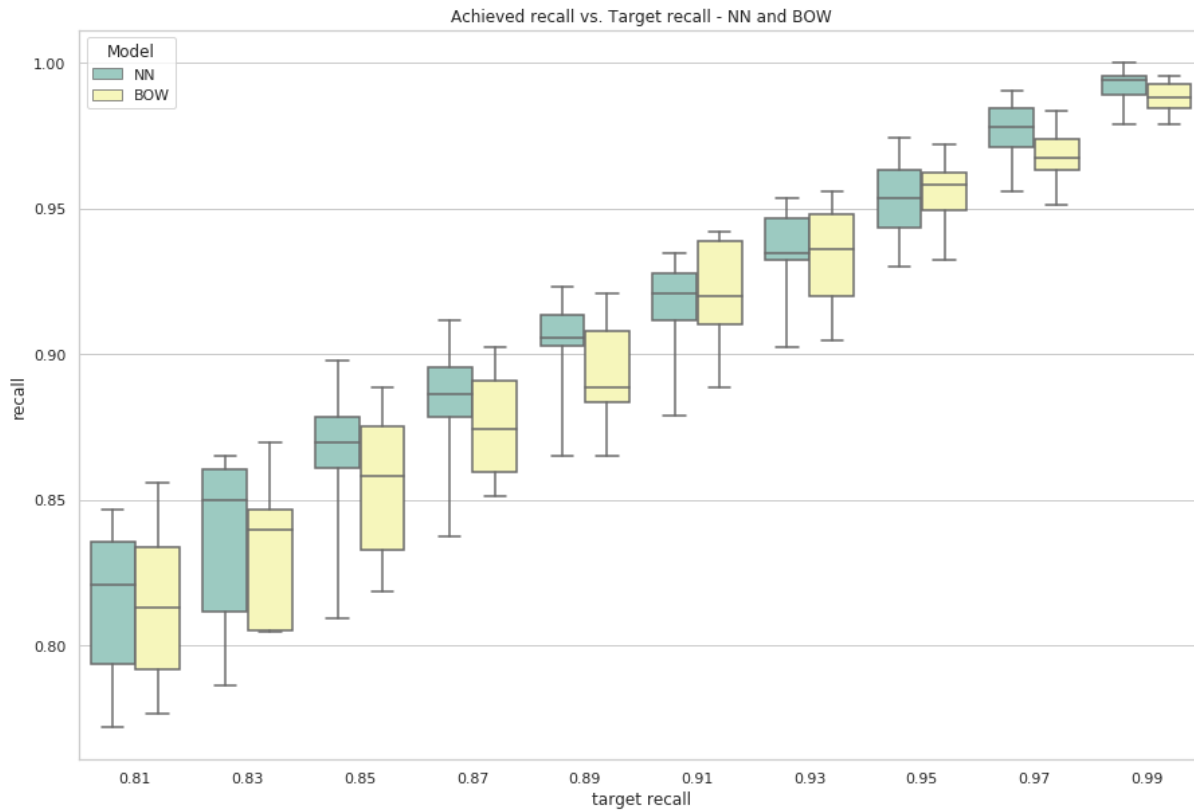

*Figure 6- Obtained vs. Target (desired) recall for NN and BOW models across 10x experiments.*

Figure 7 plots target recall against the obtained false positive rate in the test set.

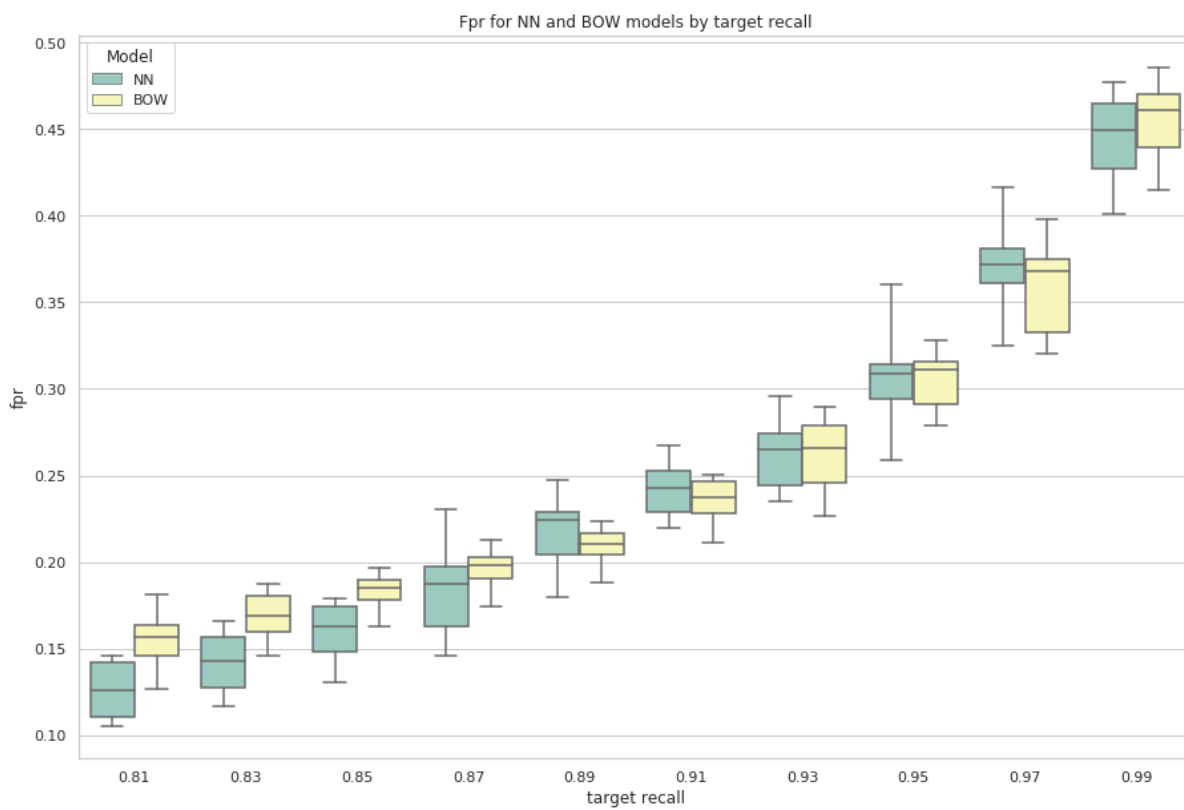

*Figure 7- Obtained FPR vs. target (desired) recall - BOW and NN models.*

In Figure 8 we compare results of NN and BOW against the strategy of combining the machine learning models with an extra rule based step (NN+BOW+Rule) from the inference pipeline. Results are shown for the 0.9-0.99 target recall range. In Table 4 we present average false positive rates from the 10 experiments.

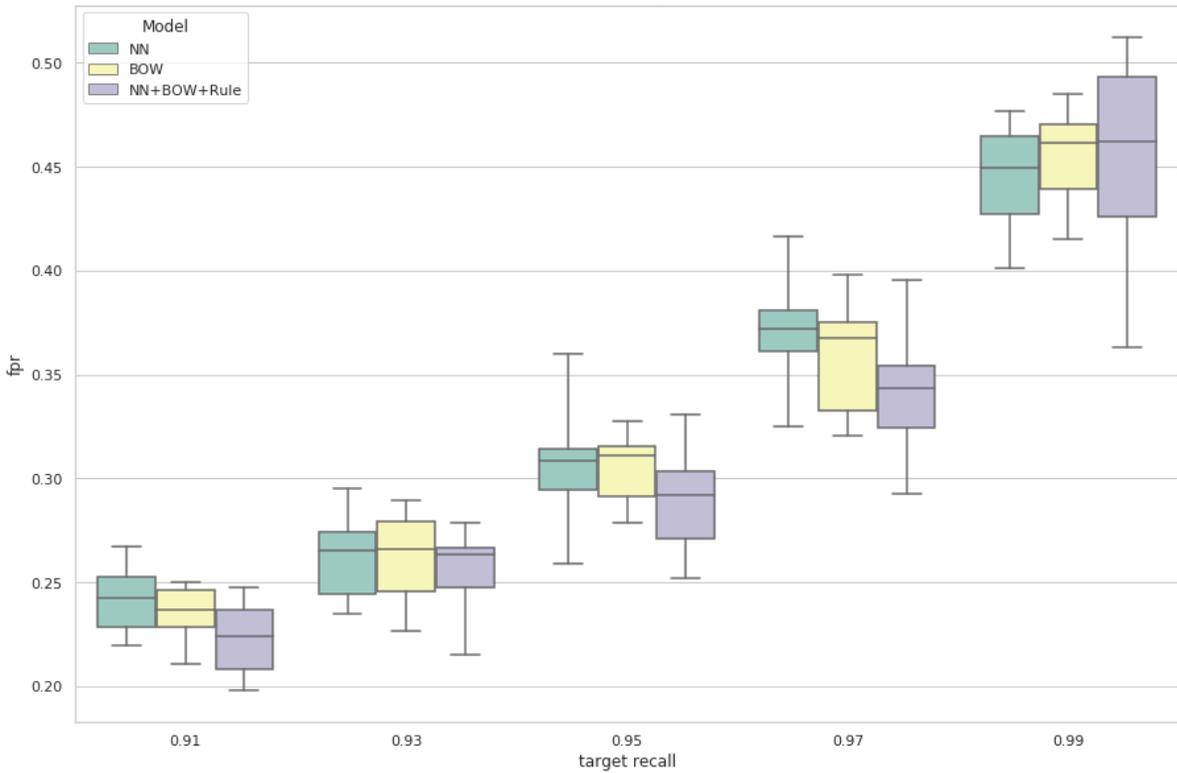

*Figure 8- Obtained FPR vs target (desired) recall - All models.*

Table 4 - Average false positive rates for given target recall by model

| Target recall | BOW | NN | NN+BOW+Rule |
|---|---|---|---|
| 0.91 | 0.23 | 0.24 | 0.22 |
| 0.93 | 0.26 | 0.26 | 0.25 |
| 0.95 | 0.30 | 0.31 | 0.29 |
| 0.97 | 0.36 | 0.37 | 0.34 |
| 0.99 | 0.45 | 0.45 | 0.45 |

## 6. Discussion

Results indicate our approach can yield productivity gains for high volume MLM workflows while maintaining high levels of recall. Taking as an example the results for a target recall of 0.99 from Table 4, the system would operate with an average of 45% false positive rate (45% of non adverse abstracts predicted as suspected adverse), thus correctly predicting 55% of non-adverse articles which could then be automatically filtered or undergo less time consuming forms of screening.

When comparing machine learning models, the NN approach provided improvements over a benchmark BOW model, but these improvements became less significant at higher levels for target recall (target recall = 0.99). Combining the BOW and NN at rule-base stage provided more robust results across all values tested except 0.99, suggesting further opportunities to experiment with combining other machine learning techniques and model ensembles.

Using hyper parameter searching to find suitable threshold values for a target recall has approximated to reasonable obtained recall values in the test set as seen in Figure 6, indicating it is a viable method for tuning the system to the particular level of risk (when lower recalls are tolerated, more productivity gains can be had). This method reflects similar studies of AI in pharmacovigilance [15] and can help support the validation needs of different applications.

With respect to risk management and controls identified in this study, we observe that public disclosures play an important part in risk mitigation: they contribute to informed decision making by end users by clearly stating the AI system's intended use, operating envelope, architecture and performance metrics. Another observation is that controls can be applied based on how the AI system is operationalized to end users - ie. the overall system it is embedded in. For instance, an interface that enables different levels of human supervision can support the progressive adoption of AI automation (requiring progressively less supervision) to suit a user's desired strategy.

In our experience the continuous involvement of pharmacovigilance experts is a critical success factor: this collaboration provided tangible benefits implemented as risk control measures across the system. Examples include refinements to the system's the intended use, data labelling and the rule-base stage in the inference pipeline. Our experience resonates with establishing multi-stakeholder development processes to improve transparency [14]. Furthermore, AI systems are often operationalized as subsystems within an end user application, as is our case[2]. Here, subject matter expertise can too provide valuable feedback on the application user experience to minimize misuse risks from AI automation.

## 7. Conclusions

In this study we presented an AI-based system for improving productivity in medical literature monitoring workflows by filtering irrelevant articles from human screeners with high levels of recall. The system is designed and built taking into particular consideration requirements and constraints from existing guidance for validated AI systems in pharmacovigilance and best practices from the AI transparency literature. We believe critically evaluating AI systems that operationalize emergent industry guidance with concrete implementations are a valuable step in driving adoption of AI technologies for pharmacovigilance and hope our study provides a useful case study to compare and contrast real world implementations of existing guidance.

---

[2] More details in: https://tinyurl.com/ytp5bf97

© 2021 biologit Ltd    15

We aim to further develop the system's inference pipeline by experimenting with recent neural architectures that have provided improvements on the state of the art on a number of NLP tasks, and by using more sophisticated robustness testing for input validation. We also wish to refine our public disclosures based on proposed guidance from [20,21] extending data disclosures and giving greater consideration to model bias and fairness evaluation.

## References

[1] Postmarketing safety reporting for human drug and biological products including vaccines - Guidance for the industry. U.S. Food and Drug Administration. (2001) https://www.fda.gov/media/72504/download. Accessed 18-Aug-2021.

[2] Guideline on good pharmacovigilance practices (GVP) Module VI – Collection, management and submission of reports of suspected adverse reactions to medicinal products (EMA/873138/2011). European Medicines Agency. July 2017. https://www.ema.europa.eu/en/documents/regulatory-procedural-guideline/guideline-good-pharmacovigilance-practices-gvp-module-vi-collection-management-submission-reports_en.pdf. Accessed 18-Aug-2021.

[3] Ghosh, R., Kempf, D., Pufko, A. et al. Automation Opportunities in Pharmacovigilance: An Industry Survey. Pharm Med 34, 7–18 (2020). DOI: 10.1007/s40290-019-00320-0

[4] Ohana, B., Baker, N., Hederman, L. Reducing screening workload in medical literature monitoring with machine learning. DIA Regulatory Science Forum. (September 2020). DOI: 10.26226/morressier.5f55fb7d6fdcfc6871991734

[5] International Society for Pharmaceutical Engineering (ISPE). Section 5—Quality Risk Management in GAMP® 5, A Risk-Based Approach to Compliant GxP Computerized Systems (2008). https://www.ispe.org.

[6] Huysentruyt, K., Kjoersvik, O., Dobracki, P. et al. Validating Intelligent Automation Systems in Pharmacovigilance: Insights from Good Manufacturing Practices. Drug Saf 44, 261–272 (2021). https://doi.org/10.1007/s40264-020-01030-2

[7] US FDA Proposed Regulatory Framework for Modifications to Artificial Intelligence/Machine Learning (AI/ML)-Based Software as a Medical Device (SaMD), Discussion Paper and Request for Feedback. (2019). https://www.fda.gov/media/122535/download. Accessed 18-Aug-2021.

[8] Git - https://git-scm.com/. Accessed 18-Aug-2021

[9] Data Version Control (DVC). GitHub Repository. https://github.com/iterative/dvc.org. Accessed 18-Aug-2021.

[10] Neumann, M., King, D., Beltagy, I., & Ammar, W. ScispaCy: Fast and Robust Models for Biomedical Natural Language Processing. Proceedings of the 18th BioNLP Workshop and Shared Task, ACL. (2019). DOI: 10.18653/v1/W19-5034

[11] Ribeiro, M., Singh S., Guestrin C. "Why Should I Trust You?": Explaining the Predictions of Any Classifier. Proceedings of the 22nd ACM SIGKDD International Conference on Knowledge Discovery and Data Mining, San Francisco, CA, USA (2016).

[12] U.S. National Library of Medicine. Cataloging: Using Medical Subject Headings (MeSH) in Cataloging. National Library of Medicine (2010). Module 3 https://www.nlm.nih.gov/tsd/cataloging/trainingcourses/mesh/intro_010.html. Accessed 18-Aug-2021.

[13] Medical Literature Monitoring (Active-substance and herbal-substance groups covered valid from June 2020). European Medicines Agency. https://www.ema.europa.eu/en/human-regulatory/post-authorisation/pharmacovigilance/medical-literature-monitoring#substances-and-medical-literature-covered-by-ema%E2%80%99s-service-section. Accessed 18-Aug-2021.

[14] Raji, I., Yang Y. "About ml: Annotation and benchmarking on understanding and transparency of machine learning lifecycles." 33rd Conference on Neural Information Processing Systems, NeurIPS (2019).

[15] Ménard, T., Barmaz, Y., Koneswarakantha, B. et al. Enabling Data-Driven Clinical Quality Assurance: Predicting Adverse Event Reporting in Clinical Trials Using Machine Learning. Drug Safety Vol. 42, 1045–1053 (2019). DOI: 10.1007/s40264-019-00831-4

[16] Arnold, Matthew, et al. "FactSheets: Increasing trust in AI services through supplier's declarations of conformity." IBM Journal of Research and Development 63.4/5 (2019): 6-1.

[17] Hochreiter, Sepp, and Jürgen Schmidhuber. "Long short-term memory." Neural computation 9, no. 8 (1997): 1735-1780.

[18] Suggested criteria for using AI/ML algorithms in GxP. Danish Medicines Agency, March 2021. https://laegemiddelstyrelsen.dk/en/licensing/supervision-and-inspection/inspection-of-authorised-pharmaceutical-companies/using-aiml-algorithms-in-gxp/. Accessed 18 Aug 2021

[19] Horizon Scanning Assessment Report – Artificial Intelligence. International Coalition of Medicines Regulatory Authorities (ICMRA). August 2021. http://www.icmra.info/drupal/sites/default/files/2021-08/horizon_scanning_report_artificial_intelligence.pdf. Accessed 20 Aug 2021.

[20] Mitchell, Margaret, Simone Wu, Andrew Zaldivar, Parker Barnes, Lucy Vasserman, Ben Hutchinson, Elena Spitzer, Inioluwa Deborah Raji, and Timnit Gebru. "Model cards for model

reporting." In Proceedings of the conference on fairness, accountability, and transparency, pp. 220-229. 2019.

[21] Gebru, Timnit, Jamie Morgenstern, Briana Vecchione, Jennifer Wortman Vaughan, Hanna Wallach, Hal Daumé III, and Kate Crawford. "Datasheets for datasets." arXiv preprint arXiv:1803.09010 (2018).